\documentclass[aps,showpacs,prl,toolkits,twocolumn]{revtex4}
\usepackage{epsfig,amssymb,amsfonts,amsmath,pstricks,psfrag,graphicx}

\makeatletter
\renewcommand{\@makefntext}[1]{\parindent=1em\noindent\hbox to 1.8em
{\hss$^{\@thefnmark}$}#1}
\renewcommand{\@footnotemark}{\hbox{\mathsurround=0pt$^{\@thefnmark}$}}

\makeatother

\begin{document}
\title{Infrared stability of quarkyonic matter with the $1/p^4$ confinement}
\author{ L. Ya. Glozman}
\affiliation{Institute for
 Physics, Theoretical Physics branch, University of Graz, Universit\"atsplatz 5,
A-8010 Graz, Austria}

\newcommand{\be}{\begin{equation}}
\newcommand{\bea}{\begin{eqnarray}}
\newcommand{\ee}{\end{equation}}
\newcommand{\eea}{\end{eqnarray}}
\newcommand{\ds}{\displaystyle}
\newcommand{\low}[1]{\raisebox{-1mm}{$#1$}}
\newcommand{\loww}[1]{\raisebox{-1.5mm}{$#1$}}
\newcommand{\lmn}{\mathop{\sim}\limits_{n\gg 1}}
\newcommand{\vpint}{\int\makebox[0mm][r]{\bf --\hspace*{0.13cm}}}
\newcommand{\too}{\mathop{\to}\limits_{N_C\to\infty}}
\newcommand{\vp}{\varphi}
\newcommand{\vx}{{\vec x}}
\newcommand{\vy}{{\vec y}}
\newcommand{\vz}{{\vec z}}
\newcommand{\vk}{{\vec k}}
\newcommand{\vq}{{\vec q}}
\newcommand{\vpp}{{\vec p}}
\newcommand{\vn}{{\vec n}}
\newcommand{\vg}{{\vec \gamma}}

\begin{abstract}
We demonstrate an exact cancellation of the infrared divergences in the
color-singlet quarkyonic matter with the $1/{\vec p}^4$ Coulomb-like
confining interaction both in the chiral symmetry broken and restored regimes.
\end{abstract}

\maketitle

{\bf 1. Introduction.}
At moderate temperatures, below the critical one, the large
$N_c$ QCD is confining up to arbitrary large baryon densities \cite{pisarski}.
This for a long time unexpected conclusion is based on a very simple
and transparent argument. In the large $N_c$ limit there are no
dynamical vacuum quark loops and hence nothing screens a confining
gluon propagator (whatever nature this propagator can be), i.e., in the strongly
interacting dense matter  gluodynamics is exactly the same as in vacuum
and hence is confining. Consequently, it is possible to identify a new
phase, called quarkyonic (which actually could consist of a few different
"subphases"). In this dense but confined phase the bulk thermodynamic
properties (like pressure) behave as $ O(N_c)$, to be contrasted with
the $O(N_c^2)$ scaling in the high temperature deconfining phase and
with the $ O(1)$ scaling in the low-temperature and density hadronic
phase.

An  interesting question arises. One typically expects that at some
critical density in the strongly interacting matter  spontaneously
broken chiral symmetry of QCD should be restored, at least due to the
Pauli blocking of the quark levels that are required for creation of a
quark condensate. Then one arrives at a paradoxical situation: at finite
density and low temperature one can expect existence of  confined but
chirally symmetric hadrons. According to the previous experience
 it was considered to be impossible. It has been 
demonstrated, however, that this is not so. It is possible to have
manifestly chirally symmetric but confined hadrons above some critical
density, at least within a model \cite{GW}. 
This model  is manifestly confining and chirally
symmetric and guarantees spontaneous breaking of chiral symmetry in the 
vacuum. The following mechanism for confining but chirally symmetric
matter at large density is observed. Even though the Lorentz-scalar
part of the quark self-energy vanishes in the chirally restored regime,
there still exists the spatial Lorentz-vector self-energy. This self-energy
is infrared-divergent and hence the single quark cannot be observed.
In a color-singlet hadron this infrared divergence cancels exactly thus
the color-singlet hadron is a finite and well-defined quantity. Consequently
in this regime the hadron mass is generated from the manifestly confining
and chirally-symmetric dynamics. If it is also a property of QCD (which is
likely, but not yet proven), then in the high density heavy ion collision 
studies
one might see not a deconfining and chirally symmetric quark matter with
single quark excitations, but rather a confining but chirally symmetric phase
with the color-singlet hadronic excitations only.

\begin{figure}
\includegraphics[width=0.3\hsize,clip=]{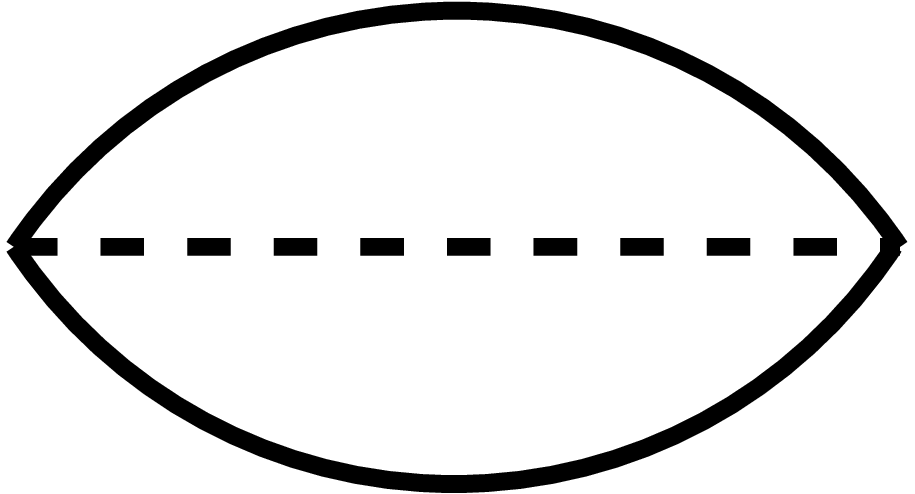}
\caption{Exchange contribution to the ground state energy.}
\end{figure}

The model relies on the linear, $ \sim 1/{\vec p}^4$, Coulomb-like confining
potential, that is indeed observed in the Coulomb-gauge QCD studies
\cite{coul} as well as in Coulomb gauge lattice simulations \cite{coullattice}.
 Certainly 
the linear confining Coulomb-like potential alone does not represent 
a complete QCD Hamiltonian
and some elements of the QCD dynamics are still missing. Nevertheless,
such a simplified model allows one to answer some principal questions
and obtain insight. 

It is 
well-known, however, that  energy of the
ground state of a many-body nonrelativistic fermion
system with the  $ \sim 1/{\vec p}^4$ interaction is divergent
already at the leading order in the interaction potential, due to graphs on
Fig. 1. Consequently, an objection was posed that the quarkyonic matter
with such an interaction  should collapse and cannot exist \cite{shuryak}.
In this report we address this issue and show that it is not so, because
the infrared divergence of a pairwise confining force is exactly canceled
by the infrared divergences of the single-quark self-energies in a color-singlet
quarkyonic matter.

 {\bf 2. Confinement and chiral symmetry properties in a vacuum}
 
 Here we briefly overview some main elements of the model.
 For the last two decades this model has been exploited many
 times with different purposes, for some of the relevant references 
 see \cite{model} and references cited therein.
 
 We work in the chiral limit and the two flavor 
 version of the model is considered. The global chiral symmetry
of the model is $U(2)_L \times U(2)_R$  because in the large $N_c$
world the axial anomaly is absent.
The only interquark interaction in our case is a linear instantaneous
Lorentz-vector  potential that has a Lorentz
structure of the Coulomb
potential:

\begin{equation} 
K^{ab}_{\mu\nu}(\vec{x}-\vec{y})=g_{\mu 0}g_{\nu 0}
\delta^{ab} V (|\vec{x}-\vec{y}|); ~~~~~
\frac{\lambda^a \lambda^a}{4}V(r) = \sigma r,
\label{KK}
\end{equation}

\noindent
where $a,b$ are color indices. The Fourier transforms of this potential
and any loop integral are infrared divergent. Hence to solve the gap
(Schwinger-Dyson equation) and the bound state equations the infrared
regularization is required. Any observable should not depend on the
infrared cut-off parameter $\mu_{IR}$ in the infrared limit, i.e.,
when this parameter is sent to 0, $\mu_{IR} \rightarrow 0$. There are
several equivalent ways to perform a regularization. Here, specifically,
we use the following regularized potential
 
\begin{equation}
V(\vec p)= \frac{8\pi\sigma}{({\vec p}^2 + \mu_{\rm IR}^2)^2}.
\label{FV} 
\end{equation}

\noindent
Then this potential in the configuration space contains the
required $\sigma r$ term,  the infrared-divergent term
$-\sigma /\mu_{IR}$ as well as terms that vanish in the infrared limit.

\begin{figure}
\includegraphics[width=1.0\hsize,clip=]{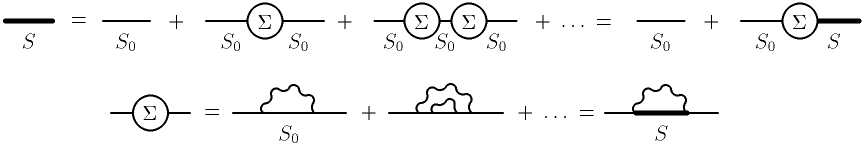}
\caption{Dressed quark Green function and the Schwinger-Dyson equation.}
\end{figure}

 Parametrizing the self-energy operator in the form

\begin{equation}
\Sigma(\vec p) =A_p +(\vec{\gamma}\cdot\hat{\vec{p}})[B_p-p],
\label{SE} 
\end{equation}

\noindent
where functions $A_p$ and $B_p$ are yet to be found, the
Schwinger-Dyson equation for the self-energy operator in 
the rainbow approximation, which is valid in the large $N_c$ limit
for the instantaneous interaction, see Fig. 2,
is reduced to the nonlinear gap equation for the chiral (Bogoliubov)
angle  $\varphi_p$,
 
 \begin{equation}
 A_p \cos \varphi_p - B_p \sin \varphi_p = 0,
 \label{gap}
 \end{equation}
 
\noindent
where
 
\begin{eqnarray}
A_p & = & \frac{1}{2}\int\frac{d^3k}{(2\pi)^3}V
(\vec{p}-\vec{k})\sin\vp_k,\quad  
\label{AB1} \\
B_p & = & p+\frac{1}{2}\int \frac{d^3k}{(2\pi)^3}\;(\hat{\vec{p}}
\cdot\hat{\vec{k}})V(\vec{p}-\vec{k})\cos\vp_k. 
\label{AB2} 
\end{eqnarray} 

The functions $A_p,B_p,$  i.e., the quark self-energy,
  are divergent in the
 infrared limit, 
 
 \begin{eqnarray}
A_p&=&\frac{\sigma}{2\mu_{\rm IR}}\sin\varphi_p+A^f_p,\nonumber \\
B_p&=&\frac{\sigma}{2\mu_{\rm IR}}\cos\varphi_p+B^f_p,
\label{iromega}
\end{eqnarray}
where $A^f_p$ and $B^f_p$ are infrared-finite functions.
This     implies that the single quark cannot be observed
 and the system is confined. However, the infrared divergence 
 cancels exactly in 
 the gap equation (\ref{gap})
so this equation can be solved directly in the infrared limit. The
chiral symmetry breaking is signaled by the nonzero chiral angle
and quark condensate
as well as by the dynamical momentum-dependent mass of quarks, $M(p)$. 
 
Given a dressed quark Green function from the gap equation, one can solve
the Bethe-Salpeter equation for mesons \cite{mesons} or variational
dynamical equations for baryons \cite{baryons}. The infrared divergences
cancel exactly in these equations for the color-singlet hadrons so the
hadron mass is a well defined and finite quantity. \footnote{A choice of
a proper infrared regularization is an important and subtle issue. As usual,
a regularization prescription is a part of a {\it definition} of the
Hamiltonian and different regularization prescriptions can imply different
physical properties of the system. Then it is the physical requirements which
determine a choice of a regularization scheme.
In our case a constraint comes from
the requirement of confinement - the physical spectrum should contain only
the color-singlet states. The regularizaton (\ref{FV}) does satisfy
this requirement.
 One could choose, however, another 
 regularization prescription, e.g., 
$\sigma r \rightarrow  \sigma r \exp(-\mu_{IR} r)$, that was also used
in the literature in the past. This prescription leads to the same
gap equation as well as to the same spectrum of the color-singlet hadrons.
However, with this prescription the single-quark energy is 
infrared-finite. This means that the spectrum of the such defined
theory contains both
the color-singlet hadrons and the colored single-quark states. Obviously 
it does not satisfy the requirement of confinement and only those
regularization prescriptions can be used that remove  single quarks
from the spectrum.}

 {\bf 3. Cancellation of the infrared divergences in quarkyonic matter}

 Before proving cancellation of the infrared divergences in  the quarkyonic
 matter it is instructive first to recall how these divergences cancel in the
 color-singlet hadrons.
 
 A (diverging) single-quark energy, i.e., a dispersive law,
  is determined by the single
 quark Green function and is given as

 \begin{equation}
\omega(p) = \sqrt( A^2_p  + B^2_p) = \frac{\sigma}{2\mu_{IR}} + \omega_f(p),
 \label{en}
 \end{equation}

\noindent
where $\omega_f(p)$ is an infrared-finite function.
Note, that the infrared divergent term is not dependent on the chiral angle.
This means that the nature of divergence is the same both in the
Wigner-Weyl and Nambu-Goldstone modes. 
 Then in the case
of a meson bound state  there  are diverging contributions from the
quark and the antiquark self-energies as well as from the quark-antiquark
interaction potential.  
  All these three contributions exactly cancel each other in the
color-singlet $\bar q q$ state,

 \begin{equation}
2\frac{\sigma}{2\mu_{IR}}  - \frac{\sigma}{\mu_{IR}} = 0. 
 \label{en}
 \end{equation}

\noindent
Consequently there are no infrared divergent contributions in the
Bethe-Salpeter equation for $\bar q q$ mesons \cite{mesons}.
An important issue is that the infrared divergent terms both in the
quark self-energies and in the  $1/{\vec p}^4$ interquark
potential do not
depend on the absolute or relative coordinates of quarks. Consequently,
the cancellation is complete and
exact, whatever distance between the quarks is. This also means that in the
infrared-divergent terms the color-dependent contributions factorize
exactly.

The relative "-" sign  between the
self-energy divergences and the divergence from the quark-antiquark
interaction kernel as well as equality
 of their absolute values
 is provided by the proper color-dependent Casimir factors.
Indeed, the color factor for the quark (or antiquark) self-energy
contribution is given by

 \begin{equation}
\langle [1]_C | \frac{\lambda^a \lambda^a}{4} | [1]_C \rangle = 4/3, 
 \label{fac1}
 \end{equation}

\noindent
 while the quark-antiquark
interaction color factor between the quark with the number "i" and
the antiquark "j" is:

 \begin{equation}
\langle \bar q_j q_i; [111]_C |-\frac{\lambda^a_i {\lambda^a_j}^*}{4}
| \bar q_j q_i; [111]_C \rangle =-4/3, 
 \label{fac2}
 \end{equation}

\noindent
where $[111]_C$ is a Young pattern for a color-singlet $SU(3)_C$ color
wave function.
Note, that the factor 4/3 is included into $\sigma$ in eq. (\ref{en}),
according to the definition (\ref{KK}).

In baryons the interquark interaction  contribution contains
 a factor

 \begin{equation}
\langle q^3; [111]_C | 
\frac{ \sum_{i < j}\lambda^a_i \lambda^a_j}{4} | q^3; [111]_C \rangle = -2. 
 \label{fac3}
 \end{equation}

\noindent
 Hence one again observes a cancellation
of the infrared divergences, because (\ref{fac3}) cancels the factor
(\ref{fac1}) multiplied by the number of quarks in a baryon
(do not forget additional factor 1/2 as seen in eq. (8) for the quark
self-energies).

Now we are in a position to demonstrate a cancellation of the infrared
divergences in a baryonic (quarkyonic) matter. We have shown
such a cancellation in each color-singlet baryon. Hence, to prove such
a cancellation in  matter we only need to show that no new infrared 
divergences appear due to  possible interaction between quarks
belonging to different color-singlet 3q subsystems.

The interquark color interaction factor in an arbitrary system of $n$ quarks
is given by the quadratic Casimir operator $C_2$ for SU(3):

 \begin{equation}
\frac{ \sum_{i < j}^n\lambda^a_i \lambda^a_j}{4}  = 
\frac{C_2}{2} - \frac{2n}{3}. 
 \label{fac4}
 \end{equation}

\noindent
However, in a color-singlet many-quark system this Casimir operator 
is exactly zero,

\begin{equation}
C_2([color ~ singlet]) = 0. 
 \label{fac5}
 \end{equation}

\noindent
This means that in a color-singlet n-quark system the color-dependent
factor is exactly the same as in a system of $n/3$ color-singlet
3q clusters, which is not true in any color-non-singlet n-quark
system.  Consequently,

 \begin{equation}
\langle [111]_C \times [111]_C \times ...| 
\frac{\sum_{i \neq j}\lambda^a_i \lambda^a_j}{4} | 
[111]_C \times [111]_C \times ...\rangle = 0, 
 \label{fac6}
 \end{equation}
 
\noindent
where $i$ and $j$ belong to different color-singlet 3q subsystems.
Cancellation is provided by the coordinate independence of
the infrared-divergent terms. A complete antisymmetry of each color-singlet
wave function in eq. (\ref{fac6}) allows one to reduce this equation
to a stronger one

 \begin{equation}
\langle [111]_C \times [111]_C \times ...| 
\frac{\lambda^a_i \lambda^a_j}{4} | 
[111]_C \times [111]_C \times ...\rangle = 0, 
 \label{fac7}
 \end{equation}
 
\noindent
for any $i$ and $j$ belonging to different color-singlet 3q subsystems.
Consequently,
no new infrared divergences appear from the interquark interaction in a
matter, beyond those which exist in each color-singlet 3q subsystem. 
In each color-singlet 3q cluster such divergences cancel by the divergences
from the quark self-energies. This proves exact cancellation of the infrared
divergences in a dense matter, in particular in its ground state.
Note that
this cancellation equally applies to both the chiral symmetry broken phase
as well as to the chirally symmetric quarkyonic matter.\footnote{In a dense
matter the infrared finite parts $A_p^f$ and $B_p^f$ of the quark self-energy
are of course not the same as in the vacuum because we have to exclude
all occupied intermediate levels with $\vec k \neq \vec p$ - 
they cannot contribute
due to Pauli blocking. The infrared-divergent part of the self-energy
is, however, exactly the same as in the vacuum, because the divergent
contribution comes from the point $\vec k = \vec p$ and the Pauli blocking
of the other levels does not affect this point.}

At this point we see a crucial difference between the 
quark matter  in a color-singlet state and a many-fermion system without
color. In the latter case the $1/{\vec p}^4$ inter-fermion interaction leads to
a divergence and such a system cannot exist. In the former case,
however, a color-dependence of such a force provides an exact cancellation
of the infrared divergences from the interquark interaction and
quark self-energies. At the same time the infrared divergences persist
in any non color-singlet system.
This is consistent with the sufficient condition of confinement, namely that
the color-singlet systems must be infrared finite, while all colored states
must be infrared divergent. 
 
{\bf Acknowledgements}
The author is indebted to E. Shuryak for rising a question, to
L. McLerran and R. Pisarski for discussions and to
P. Bicudo, F. Llanes-Estrada and A. Nefediev for correspondence. 
The author acknowledges support of the Austrian Science
Fund through the grant P19168-N16.

\end{document}